\documentstyle[twoside,fleqn,npb,epsfig]{article}
\newcommand{\AmS}{{\protect\the\textfont2
  A\kern-.1667em\lower.5ex\hbox{M}\kern-.125emS}}

\title{ What can we learn from the Caldwell plot?
\thanks{ \it  Talk given at DIS'99, Zeuthen,
Germany, April 
1999 }}

\author{ E. Gotsman \address{ School of Physics and Astronomy, Tel Aviv
University, Ramat Aviv, 69978, Israel}}
\begin{document}

\begin{abstract}

  We show that when screening corrections are included
$\frac{\partial F_{2}(x,Q^{2})}{\partial ln(Q^{2}/Q_{0}^{2})}$
is consistent with the behaviour that one expects in pQCD. Screening
corrections explain the enigma of the Caldwell plot.

\end{abstract}

\maketitle

\section{Introduction}
   The Caldwell plot \cite{Cald} of 
 $\frac{\partial F_{2}(x,Q^{2})}{\partial ln(Q^{2}/Q_{0}^{2})}$
presented at the Desy Workshop in November 1997 suprized the community.
The results appeared to indicate that we have reached a region in the 
x and $Q^{2}$ where pQCD was no longer valid.
DGLAP evolution lead us to expect that
 $\frac{\partial F_{2}(x,Q^{2})}{\partial ln(Q^{2}/Q_{0}^{2})}$
at fixed $Q^{2}$ would be a monotonic increasing function of 
$\frac{1}{x}$, whereas a superficial glance at the data suggests that the
logarithmic derivative of $F_{2}$ deviates from the expected pQCD
behaviour, and has a turnover in the region of
2 $ \leq  Q^{2} \leq $ 4 GeV$^{2}$ (see fig.1 where the ZEUS data
and the GRV'94
predictions are shown). Opinions were also voiced that the phenomena was
connected with the transition from "hard" to "soft" interactions. 

\begin{figure}[h]
\begin{center}
\epsfig{file= 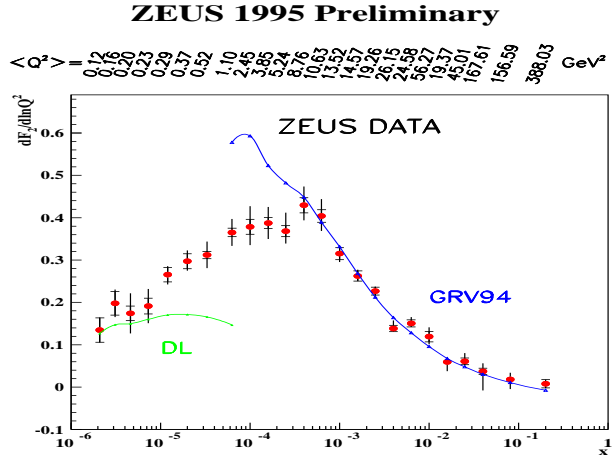,width=8cm,height=6cm}
\end{center}
\caption[fig1]{{\it ZEUS data and GRV'94 predictions for $F_{2}$ slope}}
 \end{figure}

\par Amongst the problems that one faces in attempting to comprehend the
data, is the fact that due to kinematic constraints that data is sparse,
and each point shown pertains to a different pair of values of x and
$Q^{2}$. We miss the luxury of having measurements at several different
values of x for  fixed values of $Q^{2}$, which would allow one to deduce
the
detailed behaviour of
 $\frac{\partial F_{2}(x,Q^{2})}{\partial ln(Q^{2}/Q_{0}^{2})}$.

\section{Results}
\par We show that the Caldwell plot is in agreement with the pQCD
expectations, once screening corrections (SC) (which become more important
as one goes to lower values of x and $Q^{2}$), are included. To provide a
check of our calculations, we compare with the results one derives
using the  ALLM'97  parametrization \cite{ALLM}, which we use as a "pseudo
data base". 

\begin{figure}[h]
\begin{center}
\epsfig{file= 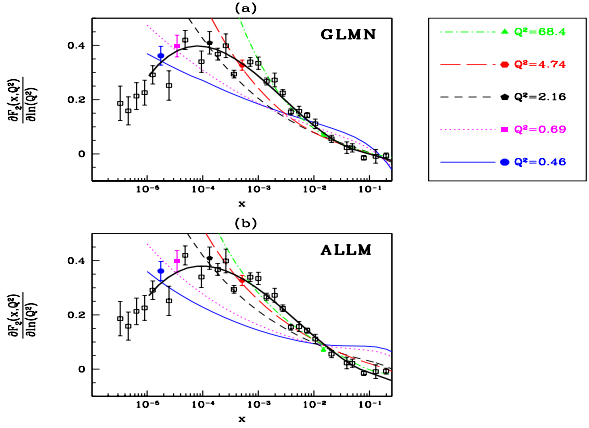,width=8cm,height=6cm}
\end{center}
\caption[fig2]{{\it The  $F_{2}$ slope  in our QCD calculation 
incorporating SC, and in the ALLM"97 parametrization. }}
 \end{figure}

\par Following the method suggested by Levin and Ryskin \cite{LR} and
Mueller \cite{M1} we calculate the SC pertaining to 
 $\frac{\partial F_{2}(x,Q^{2})}{\partial ln(Q^{2}/Q_{0}^{2})}$
for both the quark and gluon sector.  In fig.2 we show the results
as well as those of ALLM compared with the experimental results.

\begin{figure}[h]
\begin{center}
\epsfig{file= 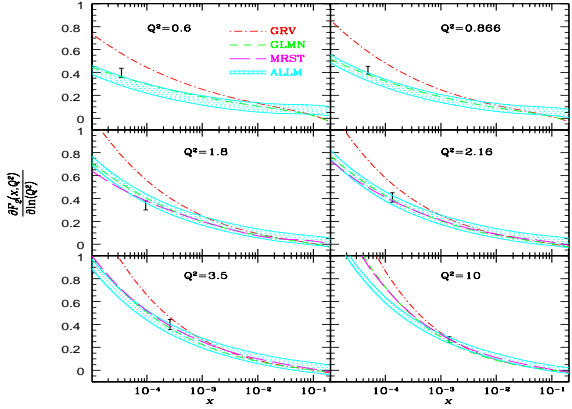 ,width=8cm,height=6cm}
\end{center}
\caption[fig3]{{\it
 $\frac{\partial F_{2}(x,Q^{2})}{\partial ln(Q^{2}/Q_{0}^{2})}$.
 In addition to the ALLM band we show a typical data point with its
error.}}
 \end{figure}

\begin{figure}[h]
\begin{center}
\epsfig{file=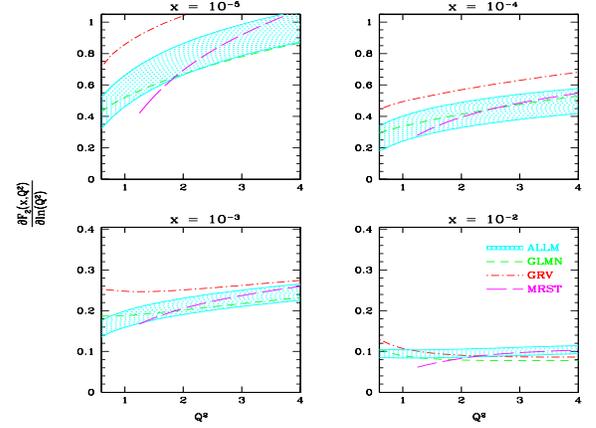,width=8cm,height=6cm}
\end{center}
\caption[fig4]{{\it
 $\frac{\partial F_{2}(x,Q^{2})}{\partial ln(Q^{2}/Q_{0}^{2})}$
at fixed x.}}
 \end{figure}

\par In fig.3 and 4  
we display our calculations for the logarithmic derivative
of $F_{2}$ after SC have been incorporated, as well as the ALLM results.
In fig.3 for fixed values of $Q^{2}$ and varying values of x, and in
fig.4 for fixed x and varying values of $Q^{2}$. In fig.4 we show our
results as well as those of ALLM compared with the experimental results.
We note that
 $\frac{\partial F_{2}(x,Q^{2})}{\partial ln(Q^{2}/Q_{0}^{2})}$
at fixed $Q^{2}$ both in our calculations and in the "psuedo data" (ALLM),
remains a $ \bf{monotonic}$ increasing function of $\frac{1}{x}$.  

From fig.4 we note that for fixed x, 
 $\frac{\partial F_{2}(x,Q^{2})}{\partial ln(Q^{2}/Q_{0}^{2})}$
decreases as $Q^{2}$ becomes smaller. The decrease becomes stronger as we
go to lower values of x. This phenomena which is due to SC adds to the
confusion in interpreting the Cadwell plot.

\section{Conclusions}
1) We have obtained a good description of 
 $\frac{\partial F_{2}(x,Q^{2})}{\partial ln(Q^{2}/Q_{0}^{2})}$
for x $ \leq $  0.1.

2) At low $Q^{2}$,
$$\frac{\partial F_{2}(x,Q^{2})}{\partial ln(Q^{2}/Q_{0}^{2})} \propto 
\:\:
Q^{2}$$ 
both in the pseudo data and in our calculations.

3) Our results suggest that there is a smooth transition between the
"soft" and "hard" processes.

4) The apparent turn over of 
 $\frac{\partial F_{2}(x,Q^{2})}{\partial ln(Q^{2}/Q_{0}^{2})}$
is an illusion, created by the experimental limitation in measuring the
logarithmic derivative of $F_{2}$ at particular correlated values of
$Q^{2}$ and x.

The detailed calculations and results that this talk was based on appear
in \cite{GLM1} and \cite{GLM2}

\section{Acknowledgements}
 I would like to thank my friends and collegues
Genya Levin and Uri Maor for an enjoyable and fruitful collaboration.

\end{document}